# Analytical Model for Electrohydrodynamic Thrust

Ravi Sankar Vaddi[1], Yifei Guan[2], Alexander Mamishev[3] and Igor Novosselov [1,4,§]

[1]*Department of Mechanical Engineering, University of Washington, Seattle, U.S.A.*

[2]*Department of Mechanical Engineering, Rice University, Houston, U.S.A.*

[3]*Department of Electrical and Computer Engineering, University of Washington, Seattle, U.S.A.*

[4]*Institute of Nano-Engineering Sciences, University of Washington, Seattle U.S.A*

**ABSTRACT**

Electrohydrodynamic (EHD) thrust is produced when ionized fluid is accelerated in an electric field due to the momentum transfer between the charged species and neutral molecules. We extend the previously reported analytical model that couples space charge, electric field, and momentum transfer to derive thrust force in 1D planar coordinates. The electric current density in the model can be expressed in the form of Mott-Gurney law. After the correction for the drag force, the EHD thrust model yields good agreement with the experimental data from several independent studies. The EHD thrust expression derived from the first principles can be used in the design of propulsion systems and can be readily implemented in the numerical simulations.

## 1. INTRODUCTION

Electrohydrodynamic (EHD) flow is the motion of electrically charged fluids under the influence of applied electric fields. EHD thrusters at their heart are simple devices consisting of two electrodes separated by an air gap and connected to a high voltage generator providing electric potential between the electrodes. When a sufficient potential is applied, the electrical breakdown of air occurs in which ions are generated near the high energy anode, known as ionization region. The ions of the same polarity as anode drift towards the ground cathode, accelerating the bulk flow by collision with the neutral molecules (in the drift region). This EHD flow propulsion phenomenon also referred to in the literature as ionic wind, has been used in many practical applications, such as convective cooling [1-3], electrostatic precipitators (ESP) [4-8], airflow control [9, 10], and as a turbulent boundary layer actuators [11]. The success of EHD technology has been limited due to the modest pressure achieved by the EHD thrusters; however, in applications where producing high pressure is not required, the EHD-driven flow can be of interest. Among the advantages of the EHD approach are the ability to operate at a small scale without moving parts, straightforward control of the system, and quiet operation. In propulsion applications, EHD converts electrical energy directly to kinetic energy, sidestepping limitation related design, and manufacturing of small moving parts, e.g., wings of micro-flyers.

The idea of using corona discharge for EHD thruster was proposed by Brown [12], who thought he has discovered an unknown phenomenon producing force and provided some explanations on the Biefeld – Brown effect. The theoretical aspect of EHD in gas was first investigated by Robinson [13], who demonstrated the ability of electrostatic blowers to generate velocities up to 4 m/s. Recently, a general analytical model was derived for planar, cylindrical, and spherical coordinates for 1D electric profiles for charge density, electric potential, and electric field strength, which can be used to calculate the 1D velocity profile [14]. The concept of electric wind associated with an EHD thrust was first demonstrated in a one-dimensional model showing that the EHD thrust is dependent on the electric pressure[15]. Christenson and Moller have developed an expression for EHD thrust and found that EHD efficiency can be related to ion mobility [16]. Moreau et al. measured EHD thrust in wire-to-cylinder corona discharge and found that the corona current $I$ is proportional to the square root of the grounded electrode diameter and to $1/d^2$, where $d$ is the spacing between two electrodes [17]. Masuyama et al. investigated both a single and dual-stage EHD thruster and showed that thrust is proportional to the square of voltage beyond the corona inception [18]. Wilson et al. investigated the use of EHD thrust for aircraft propulsion and concluded that corona discharge is not very practical for that application [19]. More recently, Gilmore showed that EHD propulsion could be viable to drive small aircraft [20], which led to the demonstration of flying fixed-wing electro-aerodynamic (EAD) aircraft [21]. Similarly, the EHD thrusters have also been proposed as a propulsion method for small-scale ionocrafts [22-24].

The EHD thrust can be modeled from the first-principles as an external force term (Coulomb force) coupled to the Navier-Stokes equations (NSE). A two-part model is required: (i) the description of the ion motion in the electrical field, and (ii) the effect of the ion drift on the neutral gas in the flow acceleration region. Several finite element and finite volume models have been developed to describe EHD velocity and pressure distributions [25, 26]. Pekker et al. first derived an ideal 1D EHD thruster model for calculating maximum thrust and thrust efficiency from the charge conservation equation and the Mott-Gurney law [27]. Mott-Gurney law describes the relation between maximum electric current density and applied voltage in semiconductors [28]. The current density was shown to vary as $J \propto \varphi(\varphi - \varphi_o)$, in agreement with the Townsend $(\varphi - I)$ relationship [29] in 1914. Since then, the form $I = C\varphi(\varphi - \varphi_o)$ has been widely adopted for corona discharge analysis [30-35], where $I$ is the corona current, $\varphi$ is the corona voltage, $\varphi_o$ is the onset corona voltage and $C$ is a fitting parameter. To physically interpret the parameter $C$, Cooperman showed that $C \propto \mu_b/L_c^2$ [36], where $\mu_b$ is the ion mobility and $L_c$ is the characteristic length scale. The thrust induced by the ions $(\varphi - T)$ relationship can be derived based on Townsend's relationship $(\varphi - I)$ [17, 18], and the maximum thrust can be defined based on Mott-Gurney law [17]. However, the scientific literature does not report an analytical model to determine the thrust induced from ions from first principles. The closest analytical model [14] couples the electrical properties and EHD driven flow was validated against the EHD velocity measurement, applied for validation of novel numerical algorithms [37], and utilized in 2D and 3D numerical simulations [25, 38, 39]; however, this model was not considered for calculations of trust force.

In this work, we extend the analytical model for EHD flow [14] to formulate the expression for EHD thrust in 1-dimensional planar coordinates. The electric current density derived from the model is presented in the form of Mott-Gurney law. The analytical model is validated against the experimental data from three independent experimental studies, including our experiments and the previously published data [17, 18].

## 2. ANALYTICAL MODEL

The analytical expressions for $(\varphi - I)$ and $(\varphi - T)$ can be derived for steady-state conditions in planar coordinates. The continuity equation for the charge density is

$$\frac{\partial \rho_e}{\partial t} + \nabla \cdot [(\mathbf{u} + \mu_b \mathbf{E})\rho_e - D_e \nabla \rho_e] = 0, \qquad (2.1)$$

where $\rho_e$ is the charge density, $\mathbf{u}$ is the velocity vector of the bulk flow, $\mathbf{E}$ is the electric field, $\mu_b$ is the ion mobility, and $D_e$ is the ion diffusivity. $D_e$ can be determined using the electrical mobility equation

$$D_e = \frac{\mu_b k_B T_e}{q}, \qquad (2.2)$$

where $k_B$ is the Boltzmann constant, $T_e$ is the absolute temperature, and $q$ is the elementary charge.
The electric field satisfies Maxwell's equation

$$\nabla \cdot \mathbf{E} = \frac{\rho_e}{\varepsilon}, \qquad (2.3)$$

where $\varepsilon$ is the permittivity, and for air, it is close to the permittivity of the space.

The ion motion is assumed to be quasi-steady since the ion drift velocity is considerably higher than the EHD-induced bulk flow. The forcing on the ions by the electric field set up by potential between the electrodes is significantly greater than the space charge diffusion, so the space charge diffusion has not been typically considered [40]. Guan et al. [25] have shown that space charge density influences the electric field lines (and thus the ion drift direction) in the vicinity of the ionization region for geometries with high angles (>45°) between the bulk flow direction and the line connecting anode and cathode in a point-to-ring geometry. In the geometry where the flow direction is aligned with electrode geometry, the space charge effect is significantly lower, and for the purpose of this derivation, is not considered. The electro-convective velocity due to external flow is negligible compared to the drift velocity [40]. The continuity equation can be reduced to

$$\nabla \cdot [\mu_b \rho_e \mathbf{E}] = 0, \tag{2.4}$$

where $\mu_b \rho_e \mathbf{E} = \mathbf{J}$ is the current flux. Combining with Eq.(2.3), the ion transport equation can be written as

$$\frac{\mu_b}{\varepsilon} \rho_e^2 - \mu_b \nabla \rho_e \nabla \varphi = 0. \tag{2.5}$$

Note that Eq.(2.5) is the same as in Sigmond [40]. Derivations for cartesian coordinates are similar to Guan [14]. Eq.(2.5) can be rearranged as

$$\nabla \varphi = \frac{\rho_e^2}{\varepsilon \nabla \rho_e}. \tag{2.6}$$

In one dimension (aligned with the flow acceleration), we have

$$\frac{d\varphi}{dx} = \frac{\rho_e^2}{\varepsilon \frac{d\rho_e}{dx}}. \tag{2.7}$$

Taking the $x$-derivative on both sides and substituting into Maxwell's equation, Eq.(2.3):

$$\frac{d^2\varphi}{dx^2} = -\frac{\rho_e}{\varepsilon} = \frac{2\rho_e \frac{d\rho_e}{dx}\left(\varepsilon\frac{d\rho_e}{dx}\right) - \rho_e^2\left(\varepsilon\frac{d^2\rho_e}{dx^2}\right)}{\left(\varepsilon\frac{d\rho_e}{dx}\right)^2}. \tag{2.8}$$

Rearranging

$$3\left(\frac{d\rho_e}{dx}\right)^2 = \rho_e\left(\varepsilon\frac{d^2\rho_e}{dx^2}\right) \tag{2.9}$$

and seeking the solution in the form

$$\rho_e = Kx^n, \tag{2.10}$$

then substituting into Eq.(2.9), the following expression is yielded

$$3n^2 x^{2(n-1)} = n(n-1)x^{2n-2}. \tag{2.11}$$

From Eq.(2.11) $n = -1/2$ and $\rho_e = Kx^{-\frac{1}{2}}$, substitute to $E = -\frac{d}{dx}\varphi$

$$-\frac{d}{dx}\varphi = -\frac{\rho_e^2}{\left(\varepsilon\frac{d\rho_e}{dx}\right)} = \frac{2K}{\varepsilon}x^{1/2} \tag{2.12}$$

$$d\varphi = -\frac{2K}{\varepsilon}x^{1/2}dx. \tag{2.13}$$

Integrating on both sides gives

$$\varphi_c - \varphi = -\frac{4K}{3\varepsilon}x^{\frac{3}{2}}. \tag{2.14}$$

The coefficient $K$ can be written as

$$K = \frac{3\varepsilon}{4x^{\frac{3}{2}}}(\varphi - \varphi_c), \tag{2.15}$$

where $\varphi$ is the applied anode potential and $\varphi_c$ is the constant potential in a corona discharge, which can be considered as potential at the $x$-location of the corona onset, or corona initiation voltage $\varphi_o$. The ion current flux between the anode and cathode is

$$J = \mu_b \rho_e E = \frac{9\mu_b \varepsilon (\varphi - \varphi_o)^2}{8x^3}. \tag{2.16}$$

The relationship in Eq.(2.16) shows that $J \propto x^{-3}$ and has a similar form to Mott-Gurney law [17, 27], i.e., $J = \frac{9\mu_b \varepsilon \varphi^2}{8d^3}$, which describes the space charge saturation limit, where $d$ is the distance between the electrodes and $\varphi$ is the applied potential. In corona discharge, the charged species are produced only after the onset potential is reached, so if $\varphi$ is replaced by $\varphi - \varphi_o$ and $x = d$, the current flux relation becomes takes the form of Mott-Gurney law.

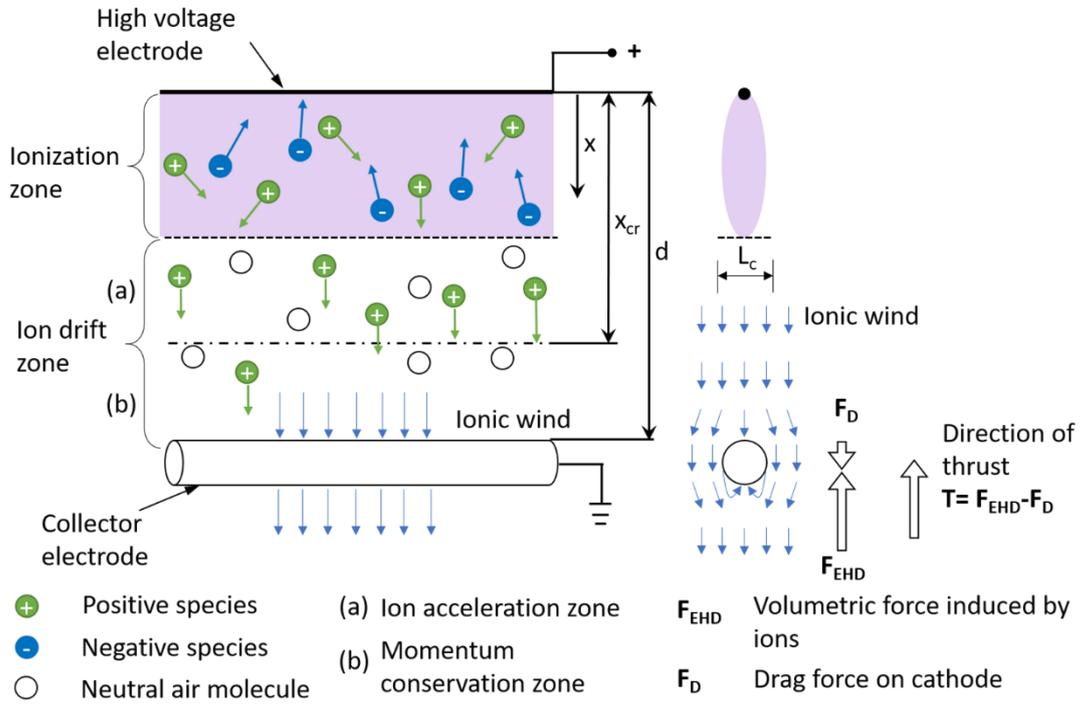

FIG. 1. Diagram of a wire-to-cylinder EHD flow. In positive corona, the negative species produced in the ionization zone recombine with positive species or the emitter (anode). The super-equilibrium positive ions drift to the collector electrode (cathode), accelerating the bulk flow. Thrust force is the resultant of the Coulombic force induced by the ions and drag force on the cathode. The conceptual representation of the EHD system includes (i) ionization region, (ii) flow acceleration region where unipolar ion motion in the gas medium acts as a body force accelerating the flow, and (iii) momentum conservation region where the electric force is balanced or overcome by viscous effects.

The EHD flow in planar wire-to-cylinder geometry can be divided into three regions: ionization zone, acceleration zone, and momentum conservation region. To define the conditions in the acceleration region, consider $x_{cr}$, which is the characteristic length scale of the flow acceleration. For wire-to-cylinder geometry, the ionization and drift regions can be approximated as an infinite plane (in the y-direction) where $x_{cr}$ is the distance from the emitter to an examination position, as shown in FIG. 1. The current flux at the location ($x_{cr}$) can be written as

$$J_{cr} = \mu_b \rho_e E = \frac{9\mu_b \varepsilon (\varphi - \varphi_o)^2}{8 x_{cr}^3}, \tag{2.17}$$

$$I = \int J_{cr} dA = J_{cr} A, \tag{2.18}$$

where $A$ is the cross-sectional area associated with ion interaction with the fluid at the location $x_{cr}$. For planar geometry (infinite length electrodes), the zone of ion interaction with the fluid can be normalized to a unit length ($L_c \times 1$). Substituting cross-section area into Eq.(2.17) gives the current expression

$$I = \frac{9\mu_b \varepsilon (\varphi - \varphi_o)^2}{8 L_c x_{cr}}. \tag{2.19}$$

To simplify, we introduce a characteristic dimension ($L_c$) that defines the ion-flow interaction region, then Eq.(2.19) can be reduced to

$$I = \frac{9\mu_b \varepsilon (\varphi - \varphi_o)^2}{8 L_c^2}. \tag{2.20}$$

This current-voltage relationship is similar to Townsend's quadratic relationship for the coaxial cylinder electrode configuration $I = C\varphi(\varphi - \varphi_o)$, where $C$ is a fitting coefficient, typically obtained from the experiments, and it is dependent on the geometry. The physical interpretation of the parameter $C$ is proposed by Cooperman for duct-type electrostatic precipitator as $C \propto \mu_b / L_c^2$, where $\mu_b$ is the ion mobility and $L_c$ is the characteristic length scale [36]. Our derivation also shows a similar physical interpretation of Townsend constant:

$$C = \frac{9\mu_b \varepsilon}{8 L_c^2}. \tag{2.21}$$

The derived ($\varphi - I$) relationship Eq.(2.19) is more general than formulations given by Townsend [29], the values of $\varphi_o$ and $L_c$ must be determined for any specific geometry. Once the ($\varphi - I$) the relationship is defined, force induced by EHD can be computed as the Coulomb force acting on the volume of fluid by the non-equilibrium concentration of ions between the anode and cathode

$$F_{EHD} = \int f dV = \int \rho_e E dV = \int_0^d \rho_e E A_L dx = \frac{Id}{\mu_b} = \frac{9\varepsilon(\varphi - \varphi_o)^2 d}{8 L_c^2}, \tag{2.22}$$

where $F_{EHD}$ is the volumetric force induced by the ions and $f$ is the force per unit volume.

Previous research [17, 18] shows the use of Townsend's current relation in Eq.(2.22) to determine the EHD force by fitting the constant $C$. However, the measured thrust does not always agree with the calculated EHD force, because the measured thrust is the result of the coulombic and drag forces. Predicted thrust force from Townsend's current voltage relationship can be 70% greater than the measured one [17, 41], likely due to losses associated with drag and the 3D field effects. The determination of drag on the cathode in a wire-to-cylinder system requires the knowledge of the velocity profile. However, the velocity measurements can be challenging near the high voltage emitter and may not be available a priori. The mean electric wind velocity $v$ and pressure $P$ can be approximated from the Bernoulli equation as

$$P = \frac{1}{2}\rho v^2, \tag{2.23}$$

where $\rho$ is the density of the fluid. The pressure gradient in the one-dimension coordinate system induced by the corona discharge can be written as

$$f = \frac{dP}{dx} \tag{2.24}$$

Combining Eq.(2.22) and Eq. (2.24), the expression for pressure can be written as

$$P = \int f \, dx = \frac{Id}{\mu_b A}. \tag{2.25}$$

The mean velocity of EHD flow can be determined from the eq (2.23)

$$v = \sqrt{\frac{2Id}{\mu_b \rho A}}. \tag{2.26}$$

The drag force due to the flow over the cathode can be calculated from the following expression

$$F_D = \frac{1}{2}\rho v^2 S C_D, \tag{2.27}$$

where $F_D$ is the drag force, $S$ is the cross-section area of the cathode and $C_D$ is the drag coefficient of the cathode. Though in the case of corona discharge, the velocity profile is not uniformed, Eq. (2.27) can be used as an approximation. Substituting Eq. (2.26) into Eq. (2.27) simplifies it further

$$F_D = \frac{Id}{\mu_b}\frac{SC_D}{A} = \theta F_{EHD}. \tag{2.28}$$

Here $\theta$ is a non-dimensionless quantity that is the ratio of the cross-section area of the cathode and corona discharge area multiplied by the drag coefficient of the cathode. The value of $\theta$ has to be less than unity and has to be determined for a specific cathode geometry. Thrust can be written as

$$T = (1-\theta)F_{EHD}. \tag{2.29}$$

The derived $(\varphi - T)$ relationship is more general than particular formulations presented in ref [17]. This formulation can be used for determining the corona current and thrust forces in planar coordinates. Unlike the thrust force formulations that use Townsend relation with fitting parameter $C$, our model captures the thrust force generated by ions including aerodynamic losses. Table 1 compares thrust characteristics derived from empirical [28, 29] data and our first-principles approach.

Table 1. Comparison of analytical expressions from state of the art and our work

|  | State of the art | Current work | Comparison |
|---|---|---|---|
| Current flux | $J = \frac{9\mu_b \varepsilon(\varphi)^2}{8d^3}$<br>Mott Gurney law [28] | $J = \frac{9\mu_b \varepsilon(\varphi - \varphi_o)^2}{8x^3}$ | Current flux at any $x$;<br>Model accounts for<br>ionization onset – $\varphi_0$ |
| Voltage - current characteristics | $I = C\varphi(\varphi - \varphi_o)$<br>Townsend relation [29] | $I = \frac{9\mu_b \varepsilon(\varphi - \varphi_o)^2}{8L_c^2}$ | Length scale $L_c$ provides<br>a physical interpretation<br>to fitting constant - $C$ |
| Voltage - force characteristics | $F_{EHD} = \frac{C\varphi(\varphi - \varphi_o)d}{\mu_b}$ | $F_{EHD} = \frac{9\varepsilon(\varphi - \varphi_o)^2 d}{8L_c^2}$ | Coulombic force is<br>computed from the first<br>principles |
| Voltage - thrust characteristics | No expression | $T = (1-\theta)F_{EHD}$ | Model computes trust and<br>accounts for aerodynamic<br>losses via parameter $\theta$ |

## 3. MODEL VALIDATION- EXPERIMENTAL SETUP

The analytical model is compared to the EHD thrust measurement in a wire-to-airfoil geometry. FIG. 2 shows the experimental setup. The emitter is a 100-micron diameter tungsten wire, the cathode is a symmetrical airfoil (NACA 0024) fabricated of a 25-micron copper sheet and has a length of 25 mm. A thruster frame of 100 mm wide was built of a polylactic acid polymer. The frame was suspended from an analytical balance Metler Toledo (AE 240) with 40 g capacity and 0.01 g resolution. The distance between the electrodes (*d*) was varied in the range of 10 to 30 mm using spacers. The thruster mass is ~ 26 g in the 10 mm spacer configuration. A high voltage power supply (Glassman, model EH30P3) was used to set the electric potential between the electrodes. The electrical connections of both electrodes were established by thin wire (100 μm in diameter) to minimize added weight. The thruster was hung from a hook on the underside of the balance using cotton strings to electrically isolate the balance and to avoid current leakage, and the thrust was measured as a reduction in weight measure by the balance. The experimental procedure is as follows: (i) the high voltage is switched off and the weight of the thruster is measured using the balance (ii) the high voltage is switched on and the difference in the balance measurements is determined, the voltage value is increased in the increments of 1 kV. The experiment was operated in the positive corona mode in a room temperature range of 22-25 °C, relative humidity of 24-26%, and ambient pressure. For each distance (*d*), the voltage was increased from 7 kV (when the thrust force becomes measurable) to ~ 29 kV (or until a spark-over occurs). To verify the measurements, each experiment was conducted five times.

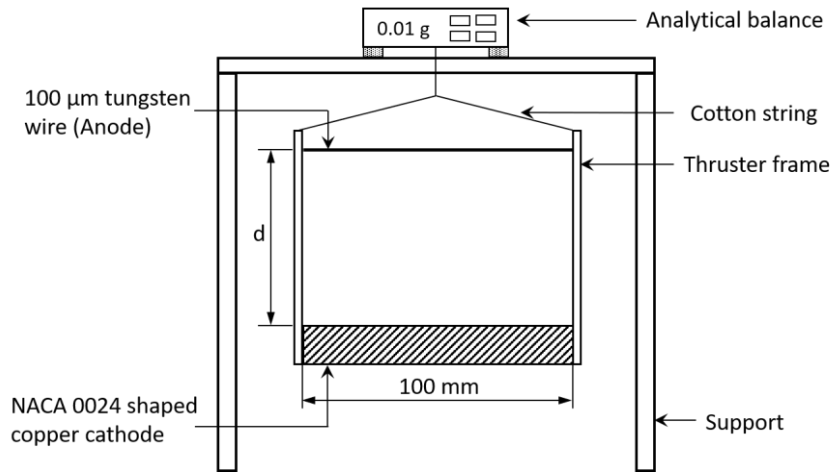

FIG. 2. Schematic of the experimental setup. A high voltage is applied between the corona wire and the ion-collecting airfoil-shaped cathode. The distance and voltage are varied in the experiments.

## 4. RESULTS AND DISCUSSION

**a) Electrode spacing effect**

The variation in distance between the electrodes has several effects (i) the strength of the E-field decreases with distance, (ii) the net thrust is proportional to the volume of the ion drift region (iii) greater electrode spacing results in higher viscous losses. FIG. 3 compares the voltage-thrust data against and the analytical solution varying $d$ = 10-30 mm and $\varphi$ = 7-29.5 kV. The relationship between the thrust and voltage is quadratic, as predicted by equation (2.22), which agrees with the trends reported in the literature for a wire to cylinder corona configuration [17, 18]. These trends can be used to estimate corona onset voltage $\varphi_o$; at this condition, the thrust is negligible. The experimental data show that higher thrust is observed at smaller gap lengths for a given voltage as the electrical field strength is greater. However, smaller gap configurations are limited due to earlier electrical breakdown (sparkover leads to a loss of thrust). The experimental thrust data is compared with two different models, (i) model without the aerodynamic drag on the cathode, see Eq. (2.22) and (ii) model with aerodynamic drag losses Eq. (2.29). As the voltage increases, the model without drag correction over-predicts the experimental thrust as the aerodynamic drag correction is greater at higher flow velocities. The analytical model with drag force correction has excellent agreement with the experimental results at lower voltages. The model agrees within ~10% at higher voltages.

In previous work [14, 40], the characteristic dimension $L_c$ was used as a fitting parameter to determine the $(\varphi - I)$ curves, which are linearly dependent on $d$. By the same logic, the best fit is obtained when $L_c = 10 + \beta(d - 10)$ when $\beta = 1$. This relationship is likely to change for other electrode configurations. The choice of $L_c$ is dependent on the drag force calculations, as well as one can see from Eq. (2.28).

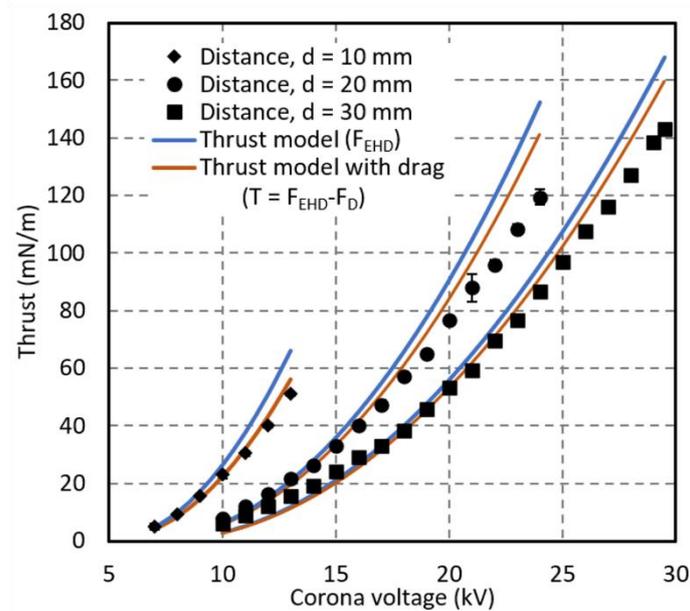

FIG. 3. Voltage-thrust relationship for varying distances between the anode and cathode for positive corona discharge. The experimental data are compared with the analytical model with and without the aerodynamic drag on the cathode.

**b) EHD thrust model comparison with previous reports**

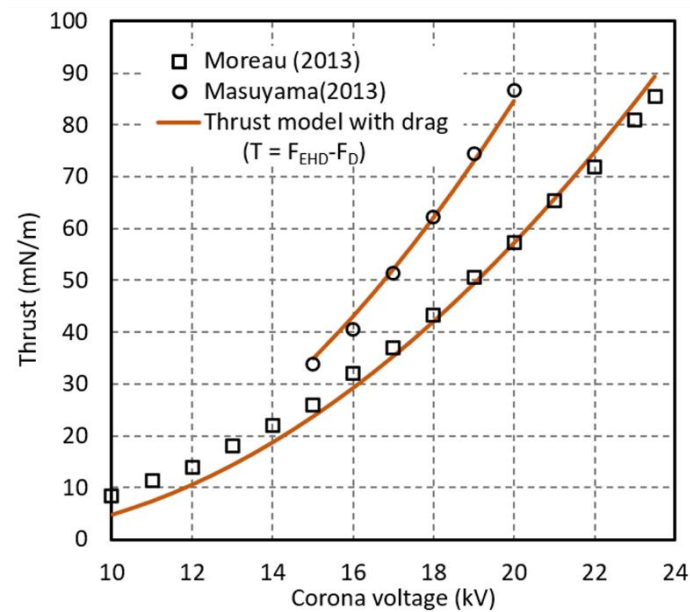

FIG. 4. Comparison of the analytical model and existing thrust data from the literature. The model with drag losses accurately predicts the thrust data for two different cases: 20 mm separation from Masuyama [18] and 30 mm separation from Moreau [17]

The model with drag correction is compared with the $(\varphi - T)$ data in wire-to-cylinder configurations, from the literature [17, 18]. FIG. 4 plots the comparison for two different electrode gaps: $d = 20$ mm [18], and $d = 30$ mm [17]. The fitting parameter $L_c = 17$ mm gives the best fit for all three studies, the model predicts the data within 10%, overpredicting the thrust at higher voltages. Though it is not apparent in our data or from Moreau et al. [17], Masuyama and Barrett [18] have observed the flattening of the ($\varphi - T$) trend at high voltages and the largest electrode gap conditions. The EHD thrust reaches saturation with the increase of potential. The presented 1-D model cannot account for this trend. At this time, we do not have an explanation for the discrepancy between the model and data at the highest voltage; however, we provide several hypotheses that may describe this behavior. (i) Nonlinear effects in the ionization region, where the increasing E-field does not produce ions at the same rate as in the lower conditions. (ii) The losses in the acceleration region due to the viscous dissipation are greater for the larger electrode gap. (iii) Space charge shielding effect, where a fraction of positive ions does not drift toward the cathode but rather are lost to surroundings (including surfaces around the experimental apparatus). This effect is likely to be enhanced as the distance between the electrodes increases. (iv) One-dimensional assumption cannot be used to describe flow, as the non-linear E-field leads to the formation of complex flow patterns. Additional investigations are required to test these hypotheses. 3D numerical modeling could be a good tool to study these effects.

## 5. CONCLUSIONS

An analytical model describing the EHD thrust is developed in 1-D coordinates and compared with data for a wire-to-airfoil and wire-to-cylinder configurations. The current density expression is analogous to Mott-Gurney law that provides the theoretical maximum of charge density between anode and cathode. The model includes a modified term to account for the corona onset voltage. The derived ($\varphi - I$) relationship has a similar form as Townsend's equation with a modified constant proportional to $\mu_b/L_c^2$. The EHD thrust force is derived from ($\varphi - I$) relationship accurately predict the thrust at lower voltages. The aerodynamic drag correction improves the agreement at the higher voltages (greater velocities). The model agrees with the experimental data from three independent studies within 10%. The limitations of the model are in predicting the thrust at the increasing voltages; these are likely the results of the simplified assumptions in the viscous losses, ionization region modeling including space charge effects, increased dimensionality of the electric field in large electrode gap geometries.

**ACKNOWLEDGMENTS**

This work was supported through an academic-industry partnership between Aerojet Rocketdyne and the University of Washington funded by the Joint Center for Aerospace Technology Innovation (JCATI) and is also based upon work supported in part by the Office of the Director of National Intelligence (ODNI), Intelligence Advanced Research Projects Activity (IARPA), via ODNI Contract 2017-17073100004. The views and conclusions contained herein are those of the author and should not be interpreted as necessarily representing the official policies or endorsements, either expressed or implied, of ODNI, IARPA, or the U.S. Government.

**NOMENCLATURE**

- $A_L$    Cross-section area of corona discharge (m²)
- $C_D$    Drag coefficient of the cathode
- $D_e$    Ion diffusivity (m²/s)
- $d$    Distance between anode and cathode (mm)
- $E$    Electric field (V/m)
- $f$    Coulomb force per unit volume (N/m³)
- $F_{EHD}$    Volumetric force induced by ions (N)
- $F_D$    Drag force (N)
- $I$    Current (A)
- $J$    Current flux [C/(s•m²)]
- $J_L$    Current flux at characteristic length scale [C/(s•m²)]
- $k_B$    Boltzmann constant
- $L_c$    Characteristic dimension
- $P$    Pressure inside a corona discharge (Pa)
- $q$    Elementary charge (C)

| | | |
|---|---|---|
| $S$ | Cross-sectional area of the cathode (m$^2$) | |
| $T$ | Thrust force induced by the ions (N) | |
| $T_e$ | Absolute temperature (K) | |
| $u$ | Velocity (m/s) | |
| $v$ | Mean electric wind velocity (m/s) | |
| $\beta$ | Scaling factor for the characteristic length | |
| $\varepsilon$ | Permittivity of air [C/(V•m)] | |
| $\mu_b$ | Ion mobility [m$^2$/(V•s)] | |
| $\rho_e$ | Charge density (C/m$^3$) | |
| $\rho$ | Density of fluid (kg/m$^3$) | |
| $\theta$ | Non-dimensionless quantity for the drag force | |
| $\varphi$ | Electric potential (V) | |
| $\varphi_o$ | Corona initiation voltage (V) | |